\documentclass{iaus}
\usepackage{graphicx}

\title[WOCS Photometric Binary Survey] 
{The WIYN Open Cluster Study Photometric Binary Survey: Initial Findings for NGC 188}

\author[Frinchaboy \& Nielsen]   
{Peter M Frinchaboy$^1$%
 \and Danielle Nielsen$^2$}

\affiliation{$^1$National Science Foundation Astronomy \& Astrophysics
Postdoctoral Fellow, \\ Univeristy of Wisconsin--Madison, 
Department of Astronomy, \\ 4506 Sterling Hall,
475 N.\ Charter Street, Madison, WI 53706, USA\\[\affilskip]
$^2$Department of Physics \& Astronomy, Colby College, \\860 Mayflower Hill Drive, Waterville, ME 04901 
\break email: frinchaboy@wisc.edu}

\pubyear{2008}
\volume{xxx}  
\pagerange{119--126}
\date{?? and in revised form ??}
\jname{Proceedings Title IAU Symposium}
\editors{A.C. Editor, B.D. Editor \& C.E. Editor, eds.}
\begin{document}

\maketitle



%
The WIYN open cluster study (WOCS) has been working to yield 
 precise magnitudes in the Johnson-Kron-Cousins $U, B, V, R, I$ system 
for all stars in the field of a selection of ``prototypical'' open clusters.
Additionally, WOCS is using radial velocities to obtain orbit solutions 
for all cluster binary stars with periods of less than 1000 days.
Recently, WOCS is being expanded to include the near-infrared $J, H, K_s$ 
(deep ground-based plus 2MASS; Skrutskie \etal\ 2006)
and mid-infrared ([3.6], [4.5], [5.8], [8.0]) photometry from {\it Spitzer}/IRAC observations.
This multi-wavelength data (0.3--8.0 $\mu$m) allows us {\it photometrically} to identify 
binaries, with mass ratios from 1.0--0.3, 
across a wide range of primary masses, especially on the faint lower main sequence (MS) 
where radial velocity surveys are prohibitive.
Here, we present work on the calibration of this technique using the cluster NGC 188, where WOCS
has also conducted an extensive search for short-period ($P \le 1000$ days) binaries 
using spectroscopy  (Geller {\it et al}., {\it in preparation})


We have combined the optical ($UBVRI$) data set of NGC 188 compiled by Stetson \etal\ (2004) 
with $JHK_s$ 2MASS data and 
deep mid-IR photometry from the Spitzer IRAC camera in HDR mode.  
For this analysis, we have restricted our sample to overlap the kinematically-studied 
WOCS sample containing MS stars 
($15.2 < V < 16.5$), with good photometry ($\sigma_{mag} < 0.1$) in all bands.  
Additionally, we selected stars with membership probability $\ge 80$\% from the 
proper motion (PM) analysis of Platais \etal\ (2003), 
resulting in a sample of 145 MS stars (Figure 1a).
The spectral energy distribution (SED) fitter by Robitaille \etal\ (2007) is used to 
fit the fluxes of 10--12 bands, converted from the observed magnitudes, to Kurucz (1979) stellar models.  
The fitted Kurucz models consist of the fluxes of single and two combined MS stars (binaries) with 
varying mass ratios using $T_{\rm eff}$, $\log(g)$, and masses from Padova isochrones 
(Girardi \etal\ 2002).  

Using our photometric technique, we find that NGC 188 has a binary fraction of 36--49\%.    
For the full sample, we found 63 ``binary'' fits out of a sample of 145 PM 
member MS stars, yielding a binary fraction of 43\%.  
However since binaries with mass ratios (MR) lower than 0.3 are difficult to distinguish from MS stars, 
we also determined the binary fraction excluding ``binary'' fits with MR $\le 0.3$ and
found 52 ``binary'' fits, as shown in Table 1.  
We have also compared our results to the spectroscopically determined binaries for NGC 188 from 
Geller {\it et al}., {\it in preparation}, which results in a spectroscopic 
binary fraction of 31--33\%.  As a result of incompleteness in the Gellar \etal\
sample, we also investigated a second smaller sub-sample ($15.2 < V < 16.0$; Figure 1a) and found 
similar binary fractions, shown in Table 1.  
Direct star-by-star comparison of the method (see Figure 1b) shows that we find roughly 2/3 of 
the spectroscopic binaries using our photometric method.  
The fact that we find a large overlap between the different
techniques is reassuring, allowing us next to explore the lower MS using this technique.

\clearpage
\begin{table}\def~{\hphantom{0}}
  \begin{center} \small
  \caption{Statistics of Binaries in NGC 188 using Photometric and Spectroscopic Techniques}
  \label{tab:kd}
  \begin{tabular}{lcccccc}\hline
 Proper Motion Member Sample & $\#$ $of$ & $Photom.$ & $Spectr.$  & $Spec\;\&\,Phot$ \\
(Prob $>80$\% and $V>13.5$)  & $Stars$   & $Binaries$& $Binaries$ & $Binaries$\\\hline
All Binaries ($V<16.5$; magenta+cyan)         & 145 & 63 (43\%) & 45 (31\%)& 31 (21\%)\\
Binaries MR $> 0.3$ ($V<16.5$)                & 145 & 52 (36\%) & 45 (31\%)& 29 (20\%)\\
All Binaries ($V<16.0$; magenta)              & 102 & 50 (49\%) & 33 (33\%)& 25 (25\%)\\
Binaries MR $> 0.3$ ($V<16.0$)                & 102 & 47 (46\%) & 33 (33\%)& 24 (24\%)\\\hline
   \end{tabular}
 \end{center}
\end{table}

\begin{figure}
\centering
\resizebox{3.94cm}{!}{\includegraphics{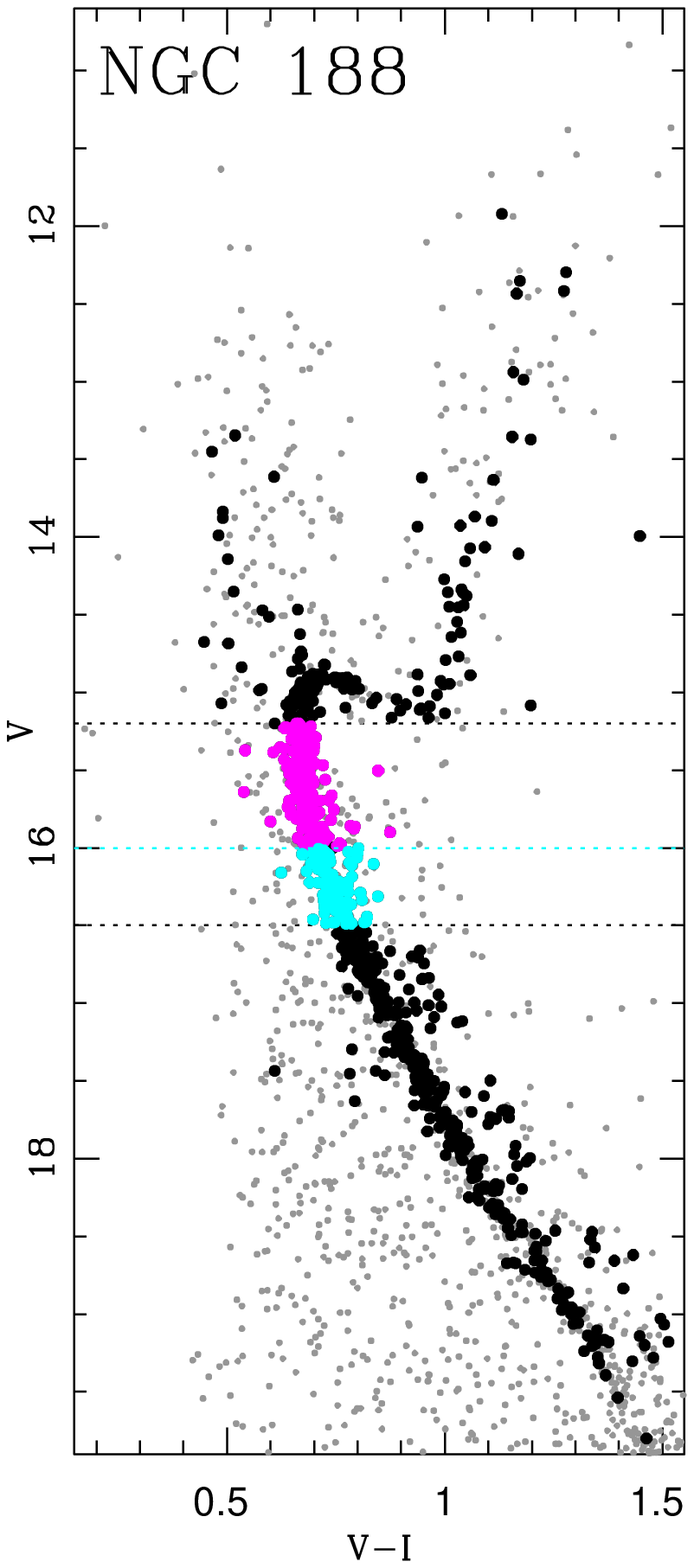} }
\resizebox{8.8cm}{!}{\includegraphics{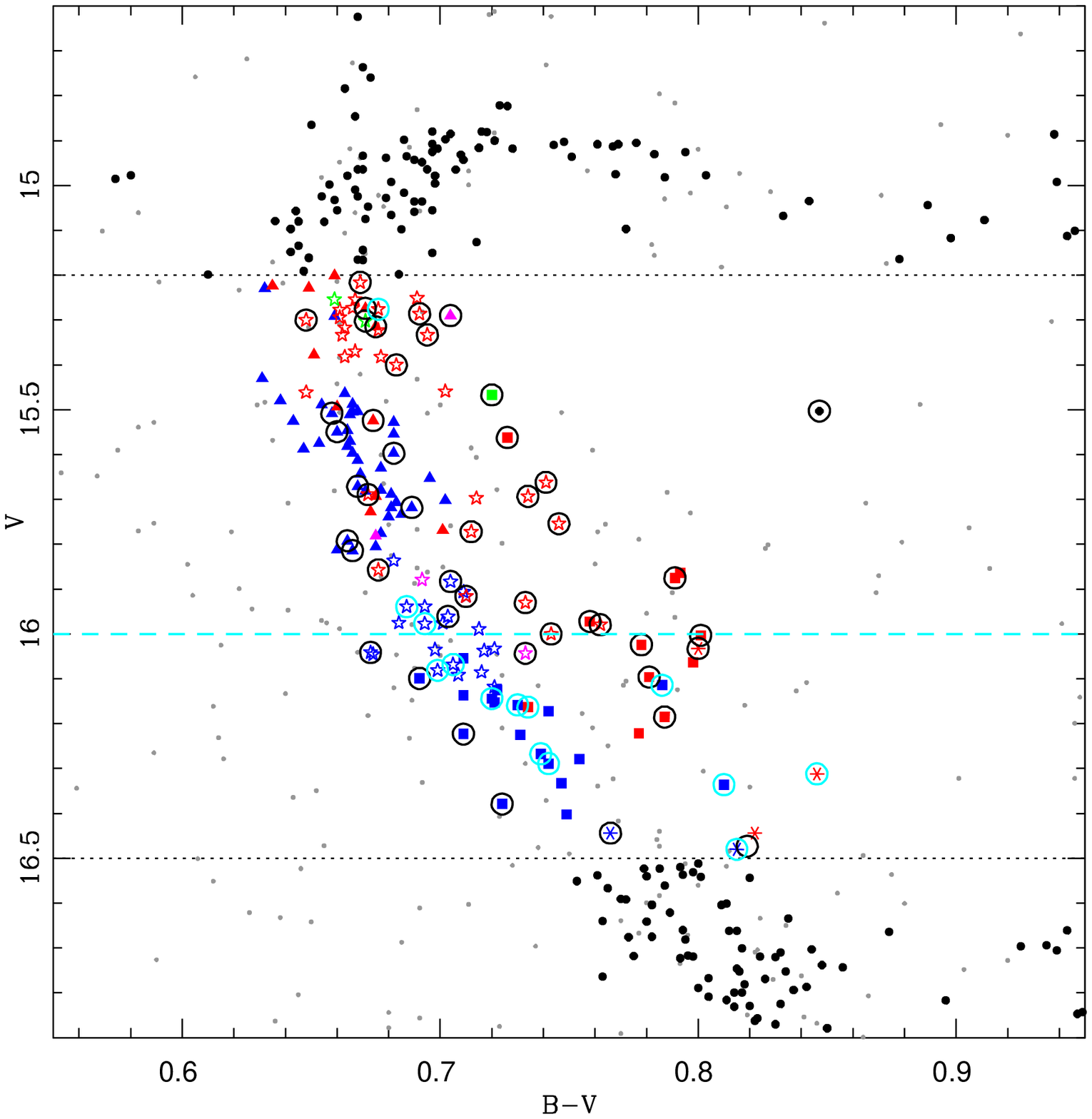} }
\vskip -0.06in
\caption[]{ Optical color-magnitude diagram (CMD) for NGC 188 using Stetson \etal\ (2004) data. 
a) Black points have membership probabilities $\ge 80$\% from Platais \etal\ (2003).  
The dotted lines delineate the sub-samples of MS stars analyzed.  
b) Red symbols are photometric binaries, blue non-binaries, with primary stars having: 
$\triangle$: $> 0.95 M_{\odot}$, $\star$: ($0.9 < M_{\odot} < 0.95$), 
$\square$: ($0.85 < M_{\odot} < 0.9$), and $*$: ($< 0.85 M_{\odot}$). 
Green symbols denote equal mass binaries.  
Black circles denote spectroscopic binaries from Geller {\it et al}., {\it in preparation}, 
while cyan circles denote stars with insufficient spectroscopic observations to determine   
if the star is a binary.}
\vskip -0.1in
\end{figure}

\begin{acknowledgments}
\vskip -0.03in
\small Any opinions,
findings, and conclusions or recommendations
expressed in this material are those of the
author(s) and do not necessarily reflect the views of the National Science
Foundation.
This project was supported by an NSF Astronomy and Astrophysics Postdoctoral Fellowship under award AST-0602221 and the NSF
REU 
program under NSF Award \# 0453442. This work is based on 
observations made with the Spitzer Space Telescope (GO-3 0800), which is operated by the Jet Propulsion Laboratory, 
California Institute of Technology under a contract with NASA. Support for this work was provided by 
NASA through an award issued by JPL/Caltech.
\end{acknowledgments}

\end{document}